\documentclass[conference]{IEEEtran}
\usepackage[utf8]{inputenc}
\usepackage{verbatim}
\usepackage{amsmath}
\usepackage{amssymb}
\usepackage{graphicx}

\makeatletter

\providecommand{\tabularnewline}{\\}

\IEEEoverridecommandlockouts
\usepackage[utf8]{inputenc}
\usepackage[T1]{fontenc}
\usepackage{cite}
\usepackage{amsfonts}\usepackage{algorithmic}
\usepackage{textcomp}
\usepackage{xcolor}
\def\BibTeX{{\rm B\kern-.05em{\sc i\kern-.025em b}\kern-.08em
    T\kern-.1667em\lower.7ex\hbox{E}\kern-.125emX}}

\makeatother

\begin{document}
\title{Multivariate Polynomial Codes for Efficient Matrix Chain Multiplication
in Distributed Systems}
\author{\IEEEauthorblockN{Jesús Gómez-Vilardebò} \IEEEauthorblockA{Centre Tecnològic de Telecomunicacions de Catalunya (CTTC/CERCA),
Barcelona, Spain, jesus.gomez@cttc.es}}
\maketitle
\begin{abstract}
We study the problem of computing matrix chain multiplications in
a distributed computing cluster. In such systems, performance is often
limited by the straggler problem, where the slowest worker dominates
the overall computation latency. To resolve this issue, several coded
computing strategies have been proposed, primarily focusing on the
simplest case: the multiplication of two matrices. These approaches
successfully alleviate the straggler effect, but they do so at the
expense of higher computational complexity and increased storage needs
at the workers. However, in many real-world applications, computations
naturally involve long chains of matrix multiplications rather than
just a single two-matrix product. Extending univariate polynomial
coding to this setting has been shown to amplify the costs---both
computation and storage overheads grow significantly, limiting scalability.
In this work, we propose two novel multivariate polynomial coding
schemes specifically designed for matrix chain multiplication in distributed
environments. Our results show that while multivariate codes introduce
additional computational cost at the workers, they can dramatically
reduce storage overhead compared to univariate extensions. This reveals
a fundamental trade-off between computation and storage efficiency,
and highlights the potential of multivariate codes as a practical
solution for large-scale distributed linear algebra tasks.
\end{abstract}


\section{Introduction}

Matrix-matrix multiplication is central to many scientific and machine
learning applications. As datasets continue to grow in scale, executing
such computations on a single server becomes impractical, and tasks
are split and executed in parallel across distributed clusters. Modern
clusters often use low-end, heterogeneous nodes prone to variability
from bottlenecks, load imbalance, or hardware faults. Such fluctuations
make overall computation time dependent on the slowest workers---a
challenge known as the straggler problem.

In this work, we address the straggler problem in distributed matrix
computations by adopting the coded computing framework introduced
in \cite{lee2017speeding,lee2017high,yu2017polynomial,yu2020straggler}.
For distributed matrix--matrix multiplication, the seminal MDS-based
formulation in \cite{lee2017speeding} introduced the notion of a
\textit{recovery threshold}, i.e., the minimum number of completed
worker computations required to reconstruct the final product. Subsequent
works proposed univariate polynomial codes \cite{yu2020straggler,dutta2019optimal}
also known as \textit{entangled polynomial} or \textit{generalized
PolyDot }codes, offering efficient decoding via polynomial interpolation
and enabling trade-offs between computation and communication costs.

Later studies \cite{amiri2019computation,kiani2018exploitation,ozfatura2020straggler}
considered assigning multiple subtasks per worker to exploit partial
computations from stragglers and improve computation latency. However,
univariate polynomial codes incur large storage requirements at workers
\cite{hasirciouglu2022JSAC,hasirciouglu2021JSAC}. To address this,
bivariate polynomial codes were introduced in \cite{hasirciouglu2021JSAC},
reducing upload costs but supporting only limited partition structures.
The recent extension in \cite{GomezITW24} generalized these codes
to arbitrary matrix partitions, achieving improved trade-offs between
computational and communication/storage overheads.

Most existing coded matrix multiplication frameworks focus on the
multiplication of two matrices. However, many practical applications---particularly
in deep learning and large-scale numerical optimization---require
matrix chain multiplication, where multiple matrices are multiplied
sequentially.%
{} The first attempt to extend coded computing to this setting appeared
in\cite[Section VI]{dutta2019optimal}, which proposed a one-shot
coding strategy based on univariate polynomial codes. %
However, this analysis was restricted to square matrices of identical
dimensions and to specific partition structures. %
A more general treatment was presented in \cite{Fan2021MatrixChain},
where a univariate polynomial coding scheme was developed for arbitrary
chain lengths and dimensions. These works showed that both the recovery
thresholds as well as the storage requirements at workers scale exponentially
with the number of matrices in the chain. 

In this work, we propose two novel multivariate polynomial coding
schemes specifically designed for distributed matrix chain multiplication.
Our analysis shows that, while the use of multivariate encodings leads
to higher recovery thresholds---and consequently increased computational
load at individual workers---it significantly reduces both storage
and communication overheads. Notably, the storage cost at each worker
no longer grows exponentially with the length of the matrix chain.

The resulting trade-off uncovers a new balance between computational
complexity and storage efficiency, establishing multivariate codes
as an effective and scalable framework for distributed implementations
of matrix-chain operations. 

The remainder of the paper is organized as follows. In Section \ref{sec:System-Model},
the system model and problem formulation is presented. Section \ref{sec:Univariate-Coded-Computing}
reviews univariate polynomial coding scheme. Next, Section \ref{sec:Multi-variate-Coded-Computing}
introduces the proposed multivariate polynomial codes. Numerical results
are provided in Section \ref{sec:Numerical-Evaluation}. Finally,
conclusions are drawn in Section \ref{sec:Numerical-Evaluation}.

\section{System Model\label{sec:System-Model}}

\begin{figure}
\begin{centering}
\includegraphics[scale=0.11]{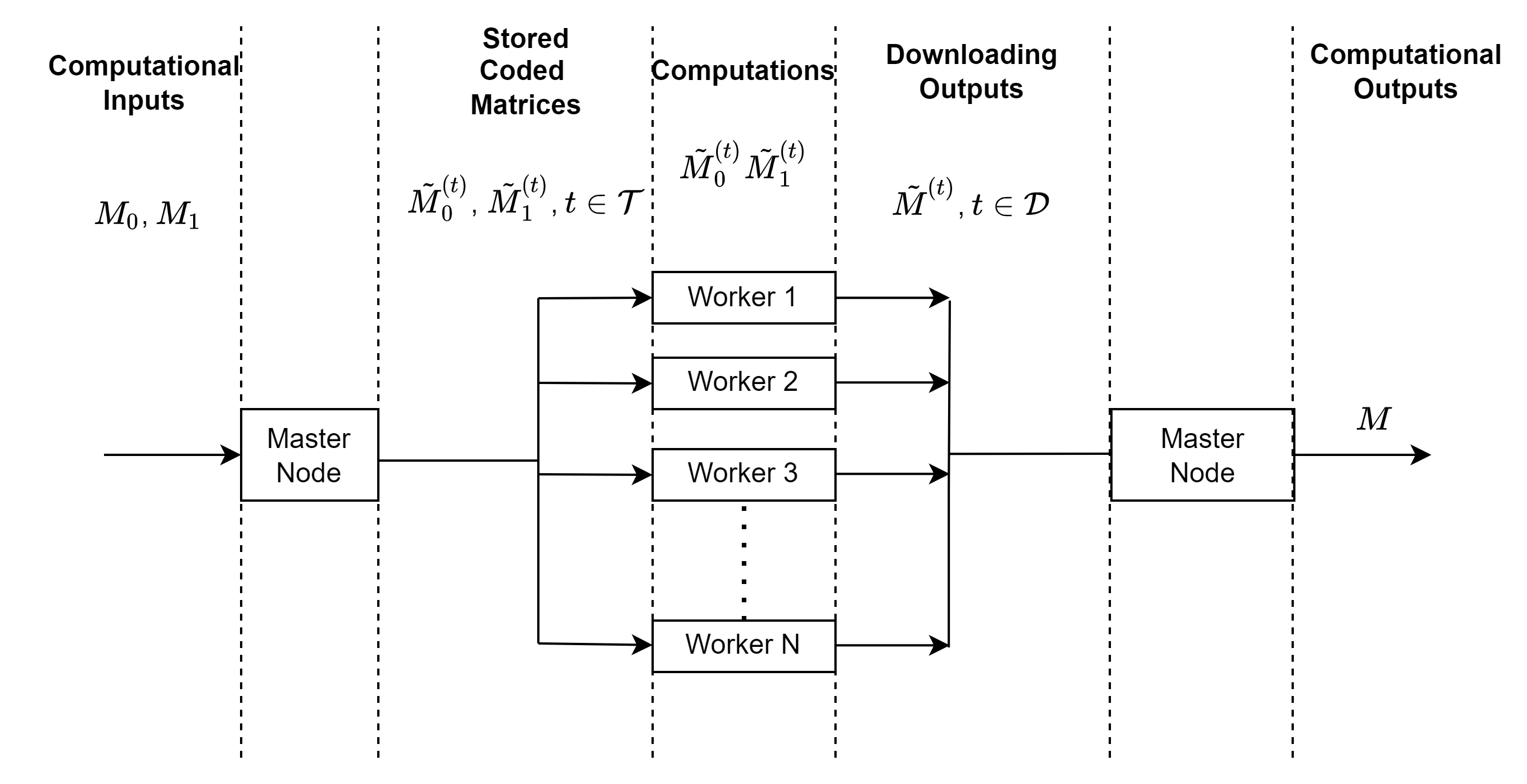}
\par\end{centering}
\caption{Distributed computational system. \label{fig:Computational-system}}
\vspace{-0.5cm}
\end{figure}

Denote the integer set $[p]:=[0,\dots,p-1]$. Consider the problem
of outsourcing the task of multiplying a chain of $m$ large matrices
$M=\prod_{i=0}^{m-1}M_{i}$, with $M_{i}\in\mathbb{F}^{r_{i}\times r_{i+1}}$,
$i\in[m],$ from a master server to $N$ distributed workers running
in parallel, see Figure \ref{fig:Computational-system} $(m=2$).
In order to distribute the total computation work, first matrices
are partitioned into matrix blocks. We consider a very general partition
scheme, in which matrix $M_{i}$ is spitted into $p_{i}$ partitions
vertically and $p_{i+1}$ partitions horizontally, resulting in $p_{i}p_{i+1}$
matrix blocks $M_{i}^{b_{i},b_{i+1}}\in\mathbb{F}^{\frac{r_{i}}{p_{i}}\times\frac{r_{i+1}}{p_{i+1}}}$
for $b_{i}\in\left[p_{i}\right]$ and $b_{i+1}\in\left[p_{i+1}\right]$,
i.e.,

\[
M_{i}=\left[\begin{array}{ccc}
M_{i}^{0,0} & \cdots & M_{i}^{0,p_{i+1}-1}\\
\vdots & \ddots & \vdots\\
M_{i}^{p_{i}-1,0} & \cdots & M_{i}^{p_{i}-1,p_{i+1}-1}
\end{array}\right].
\]
We refer to the ($m+1$)-tuple $\left(p_{0},\dots,p_{m}\right)$ as
the \textit{partition scheme}. The result of the matrix chain multiplication,
$M$, can then be decomposed into $p_{0}p_{m}$ matrix blocks 
\[
M=\left[\begin{array}{ccc}
M^{0,0} & \cdots & M_{i}^{0,p_{m}-1}\\
\vdots & \ddots & \vdots\\
M^{p_{0}-1,0} & \cdots & M_{i}^{p_{0}-1,p_{m}-1}
\end{array}\right]
\]
where matrix blocks $M^{n_{0},n_{m}}$ for $n_{0}\in\left[p_{0}\right]$,
$n_{m}\in\left[p_{m}\right]$ can be obtained, as a function of the
input matrix blocks, as

\begin{equation}
M^{n_{0},n_{m}}=\sum_{\mathbf{n}\in\mathcal{N}_{1}^{m-1}}M^{(n_{0},\mathbf{n},n_{m})}.\label{eq: product matrix block decomposition}
\end{equation}
Here $M^{(n_{0},\mathbf{n},n_{m})}$ represents the matrix block chain
product
\begin{equation}
M^{(n_{0},\mathbf{n},n_{m})}=M_{0}^{n_{0},n_{1}}M_{1}^{n_{1},n_{2}}\dots M_{m-1}^{n_{m-1},n_{m}},\label{eq:block matrix chain multiplication}
\end{equation}
the index vector is $\mathbf{n}=[n_{1},\dots,n_{m-1}]$, and the summation
is over the product set 
\[
\mathcal{N}_{1}^{m-1}=[p_{1}]\times[p_{2}]\times\dots\times[p_{m-1}].
\]
The total number of matrix chain multiplications of the form (\ref{eq:block matrix chain multiplication})
that are needed to recover the original matrix chain multiplication
is thus $K=\prod_{i=0}^{m}p_{i}$. We refer to $K$ as the \textit{partition
level. }

\textit{Coded distributed computing} strategies for matrix--matrix
multiplication or matrix chain multiplication generally proceed as
follows. First, we define as set of subtasks $\mathcal{T}$ with $\left|\mathcal{T}\right|$
elements, to be assigned to the workers. For each subtask $t\in\mathcal{T}$,
the master node generates coded matrix blocks $\tilde{M}_{i}^{(t)}\in\mathbb{F}^{\frac{r_{i}}{p_{i}}\times\frac{r_{i+1}}{p_{i+1}}}$
for all matrices in the chain $i\in[m].$ These coded blocks are obtained
by linearly encoding the original blocks of $M_{i}$. The coded matrix
blocks corresponding to different subtasks are then distributed across
the $N$ workers. Each worker $w$ is assigned a subset of subtasks
$\mathcal{T}_{w}\subseteq\mathcal{T}$ and computes, for every $t\in\mathcal{T}_{w}$
, the block matrix product chain $\tilde{M}^{(t)}=\prod_{i=0}^{m-1}\tilde{M}_{i}^{(t)}\in\mathbb{F}^{\frac{r_{0}}{p_{0}}\times\frac{r_{m}}{p_{m}}}$
. As soon as a worker finishes a subtask, it returns the result to
the master node and process to compute the next one. Finally, once
the master has collected a sufficient number of computations, equal
to the \textit{recovery threshold}, $R_{th}$, which depends on the
chosen coding strategy, it can decode the complete result of the matrix
chain multiplication. 

Notice that, unlike most works in the literature that restrict the
analysis to a single subtask per worker, we allow each worker to handle
multiple subtasks. This removes the unrealistic assumption that the
number of workers must match the number of subtasks, enables the exploitation
of partial computations (since a worker’s total workload is naturally
divided into subtasks), and supports heterogeneous environments, where
workers with different computational capabilities can be assigned
different numbers of subtasks. Furthermore, as in \cite{Fan2021MatrixChain}
we restrict our analysis to coded distributed computing strategies
that require only a single communication round. That is only the original
matrices are encoded---none of the intermediate products in the chain---and
decoding is performed in a single stage.

The recovery threshold, $R_{th},$ of a coding strategy serves as
a key performance metric. It reflects the computational complexity
of the strategy and, together with the statistics of worker finishing
times, is sufficient to estimate the overall computation latency.
Another important parameter is the storage capacity required at the
workers, or equivalently, the uploading communication cost. This corresponds
to the total amount of data transmitted from the master to the workers
and stored at them in the form of coded matrix blocks. While coded
computing reduces computation latency, it generally incurs additional
storage and computation overheads. To evaluate these overheads across
different strategies, we use as a reference the baseline case of outsourcing
the entire computation to a single powerful server. Below, we define
these terms in more detail.

We define the \textit{computation complexity} as the number of element-wise
multiplications required to obtain the result. In the single-server
setting, matrix chain multiplication involves $C^{\text{SS}}=\prod_{i=0}^{m}r_{i}$
element-wise multiplications\footnote{More precisely, the standard cubic-time algorithm achieves a complexity
of $\mathcal{O}\left(\prod r_{i}\right)$ in the general case, whit
possible reductions under specific conditions.}. In the coded computing setting, each matrix block chain multiplication
requires $\prod_{i=0}^{m}\frac{r_{i}}{p_{i}}$ element-wise multiplications.
With a recovery threshold, $R_{th}$, the total complexity becomes
$C^{\text{CC}}=R_{th}\prod_{i=0}^{m}\frac{r_{i}}{p_{i}}$. We then
define the complexity overhead, $\delta$, as
\begin{eqnarray*}
\delta+1 & = & \frac{C^{\text{CC}}}{C^{\text{SS}}}=\frac{R_{th}}{K}.
\end{eqnarray*}

The \textit{storage capacity threshold} is defined as the minimum
worker storage required to produce enough computations for recovery.
In practice, additional storage is often used to let faster workers
complete more subtasks, but we use this definition strictly for comparing
coding strategies. Storage is measured in units of original matrix
entries. In the single-server setting, storing the complete matrix
$M_{i}$ requires $S_{i}^{\text{SS}}=r_{i}r_{i+1}$ storage units.
In the coded computing setting, each coded block $\tilde{M}_{i}^{(t)}$
requires $\frac{r_{i}}{p_{i}}\frac{r_{i+1}}{p_{i+1}}$ storage units.
Let $S_{th,i}$ denote the number of coded blocks of $M_{i}$ that
must be stored (across all workers) to obtain exactly $R_{th}$, computations.
The corresponding storage capacity for $M_{i}$ is $S_{i}^{\text{CC}}=S_{th,i}\frac{r_{i}r_{i+1}}{p_{i}p_{i+1}}.$
The storage overhead associated with $M_{i},$ for $i\in[m]$, is
thus
\begin{eqnarray*}
\delta_{s,i}+1 & = & \frac{S_{i}^{\text{CC}}}{S_{i}^{\text{SS}}}=\frac{S_{th,i}}{p_{i}p_{i+1}}.
\end{eqnarray*}

\section{Univariate Coded Computing\label{sec:Univariate-Coded-Computing}}

We begin by reviewing the computation complexity and storage requirements
of the univariate polynomial scheme presented in \cite{Fan2021MatrixChain}
for matrix chain multiplications. For details on the coding strategy
itself, we refer the reader to that work. According to \cite[Theorem 1]{Fan2021MatrixChain},
the univariate polynomial scheme achieves a recovery threshold of
\[
R_{th}^{(\text{UV})}=\prod_{j=0}^{m}p_{j}+\prod_{j=1}^{m-1}p_{j}-1.
\]
For each subtask, workers store one coded matrix block per input matrix.
Hence, to obtain exactly $R_{th}^{(\text{UV})}$ computations, the
total number of coded blocks from each matrix $M_{i}$ that must be
stored across all workers is

\[
S_{th,i}^{(\text{UV})}=R_{th}^{(\text{UV})}.
\]
To build intuition, consider the representative case where all the
matrices are partitioned equally, i.e. $p_{i}=p$, for all $i\in\left\{ 0,1,\dots,m\right\} $.
In this setting,
\[
R_{th}^{(\text{UV})}=S_{th,i}^{(\text{UV})}=p^{m+1}+p^{m-1}-1.
\]
Relative to the single-server setting baseline, the computation complexity
overhead and storage overhead become
\begin{eqnarray*}
\delta^{(\text{UV})} & = & p^{-2}-p^{-(m+1)}\sim\mathcal{O}\left(p^{-2}\right),\\
\delta_{s}^{(\text{UV})} & = & p^{m-1}+p^{m-3}-p^{-2}-1\sim\mathcal{O}\left(p^{m-1}\right)
\end{eqnarray*}
Thus, compared to the single-server setting, the computation overhead
of the univariate scheme vanishes as $p^{-2}$ when $m\geq2$. However,
the storage requirement grows rapidly with $p^{m-1}$. This steep
increase in storage (or, equivalently, upload communication cost)
represents a significant limitation for the practical deployment of
univariate coded computing in matrix chain multiplication. This motivates
the exploration of alternative strategies with lower storage and communication
costs, such as multivariate polynomial codes, which we study in the
next section.

\section{Multi-variate Coded Computing \label{sec:Multi-variate-Coded-Computing}}

In this section, we present two coding schemes for matrix chain multiplication
based on multi-variate polynomials. The objective is to reduce storage
and upload communication costs compared to univariate polynomial coding
schemes, at the expense of increased computation complexity.

\subsection{Multi-variate Scheme 1}

In this scheme, the block matrices of each of the $m$ input matrices
are encoded using a univariate polynomial, each in a different variable,
$x_{i}$, $i\in[m]$. Specifically, blocks of input matrix $M_{i}$
are encoded as

\[
\tilde{M_{i}}^{(\text{MV}_{1})}(x_{i})=\sum_{b_{i}=0}^{p_{i}-1}\sum_{b_{i+1}^{\prime}=0}^{p_{i+1}-1}M_{i}^{(b_{i},b_{i+1}^{\prime})}x_{i}^{p_{i+1}b_{i}+b_{i+1}^{\prime}}.
\]
{} Let $\mathbf{x}=\left(x_{0},\dots,x_{m-1}\right)$. The product polynomial
is then
\[
\tilde{M}^{(\text{MV}_{1})}(\mathbf{x})=\prod_{i=0}^{m-1}\tilde{M_{i}}^{(\text{MV}_{1})}(x_{i}).
\]
Expanding the product yields%
\begin{align}
\tilde{M}^{(\text{MV}_{1})}(\mathbf{x}) & =\sum_{\substack{b_{0}\in[p_{0}]\\
b_{m}^{\prime}\in[p_{m}]
}
}\sum_{\substack{\mathbf{b}\in\mathcal{N}_{1,m-1}\\
\mathbf{b}^{\prime}\in\mathcal{N}_{1,m-1}
}
}M^{(b_{0},\mathbf{b},\mathbf{b}^{\prime},b_{m}^{\prime})}\nonumber \\
 & \quad\times\prod_{i=0}^{m-1}x_{i}^{p_{i+1}b_{i}+b_{i+1}^{\prime}}.\label{eq:product polynomial MV1}
\end{align}
with the shorthand $M^{(b_{0},\mathbf{b},\mathbf{b}^{\prime},b_{m}^{\prime})}=M_{0}^{(b_{0},b_{1}^{\prime})}M_{1}^{(b_{1},b_{2}^{\prime})}\cdots M_{m-1}^{(b_{m-1},b_{m}^{\prime})}$.
Recall the decomposition of the product matrix blocks as a function
of the input matrix blocks in (\ref{eq: product matrix block decomposition}).
Observe that for any index vector $\mathbf{n}\in\mathcal{N}_{1}^{m-1},$
the matrix block chain product $M^{(n_{0},\mathbf{n},n_{m})}=M^{(n_{0},\mathbf{n},\mathbf{n},n_{m})}$
can be obtained as the coefficient of the product polynomial in (\ref{eq:product polynomial MV1})
associated to the monomial $\prod_{i=0}^{m-1}x_{i}^{p_{i+1}n_{i}+n_{i+1}}$
at the summation index $b_{0}=n_{0}$, $b_{m}=n_{m}$ and $\mathbf{b}=\mathbf{b}^{\prime}=\mathbf{n}$.

\textbf{Recovery threshold:} The polynomial degree in each variable
is $\text{deg}\left(x_{i}\right)=p_{i+1}p_{i}-1$. Hence, multivariate
interpolation requires
\begin{eqnarray}
R_{th}^{(\text{MV}_{1})} & = & \prod_{i=0}^{m-1}p_{i}p_{i+1}=p_{0}p_{m}\prod_{i=1}^{m-1}p_{i}^{2}\label{eq:Rth - MV1}
\end{eqnarray}
evaluations of $\tilde{M}^{(\text{MV}_{1})}(\mathbf{x})$. Observe
that this threshold is, in general, considerably higher than in the
univariate case, which means more computations are required, for a
given partition scheme. In contrast, significantly reduced storage
and upload communication cost are required as we show next.

\textbf{Storage capacity:} Consider first the case in which workers
share a common memory. Notice, first that under this assumption, the
storage requirements of the univariate polynomial coding strategy
remains unchanged. Let us denote by S$_{th,i}^{(\text{MV}_{1},\text{S})}$
the number of code blocks stored in the case of a shared memory. In
this case, workers need to store $S_{th,i}^{(\text{MV}_{1},\text{C})}=p_{i}p_{i+1}$
coded blocks for every input matrix $\tilde{M_{i}}^{(\text{MV}_{1})}(x_{i})$,
where each block is evaluated at a distinct point, i.e. $x_{i}\in\mathcal{X}_{i}$
with $\left|\mathcal{X}_{i}\right|=p_{i}p_{i+1}$. This matches exactly
the number of uncoded partitions required in the single-server setting.
Subsequently, by computing the product over all possible combinations
of coded block matrices --which together sum to $R_{th}^{\text{(MV}_{1})}$
-- workers can obtain the required number of computations for recovery.
Observe that $S_{th,i}^{(\text{MV}_{1},\text{S})}$, compared to its
univariate counterpart $S_{th,i}^{(\text{UV})}$, yields a significant
reduction in worker storage requirements. Now consider the case of
$N$ workers each equipped with a dedicated memory. Similarly, denote
by S$_{th,i}^{(\text{MV}_{1},\text{D})}$ the number of code blocks
stored in the case of a dedicated memory. For simplicity, we assume
homogeneous storage capacity across workers, noting that the extension
to the heterogeneous case is straightforward. Each worker is assigned
a fraction $s_{i}$ (with $0<r_{i}\leq1)$ of the $p_{i}p_{i+1}$
coded blocks of $M_{i}$, where $s_{i}$ is chosen such that $s_{i}p_{i}p_{i+1}$
is an integer. With $N$ workers in total, the aggregate number of
coded blocks stored across the system is $S_{th,i}^{(\text{MV}_{1},\text{D})}=Ns_{i}p_{i}p_{i+1}.$
By computing all the matrix coded block chain products for all the
possible combinations of coded block matrices locally available at
each worker, each worker is able to obtain up to $\prod_{j=0}^{m-1}s_{i}p_{i}p_{i+1}$
evaluations of the product polynomial. Thus, to ensure recovery, it
is necessary that
\begin{equation}
N\prod_{i=0}^{m-1}s_{i}p_{i}p_{i+1}\geq R_{th}^{\text{(MV}_{1})}
\end{equation}
or, equivalently $N\prod_{i=0}^{m-1}s_{i}\geq1$. 

Again to build intuition, choose $s_{i}=N^{-\frac{1}{m}}$ for all
$i\in[m]$ and consider the representative case where all the matrices
are partitioned equally, i.e. $p_{i}=p$, for all $i$. In this setting,
\[
R_{th}^{(\text{MV}_{1})}=p^{2m}\text{, }S_{th,i}^{(\text{MV}_{1},\text{S})}=p^{2}\text{, }S_{th,i}^{(\text{MV}_{1},\text{D})}=N^{\frac{m-1}{m}}p^{2}
\]
Thus, relative to the single-server setting
\begin{eqnarray*}
\delta^{(\text{MV}_{1})} & = & p^{m-1}-1\sim\mathcal{O}\left(p^{m-1}\right),\\
\delta_{s}^{(\text{MV}_{1},\text{S})} & = & 0\text{, }\delta_{s}^{\text{(MV}_{1},\text{D})}=N^{\frac{m-1}{m}}-1.
\end{eqnarray*}
Compared to the univariate scheme, the storage overhead are much lower
(constant or scaling only with $N$), but computation overhead grows
as $\mathcal{O}\left(p^{m-1}\right)$.

\subsection{Multi-variate Scheme 2}

In this scheme, the blocks of each matrix $M_{i}$ are encoded into
a bivariate polynomial:
\[
\tilde{M_{i}}^{(\text{MV}_{2})}(x_{i},x_{i+1})=\sum_{b_{i}=0}^{p_{i}-1}\sum_{b_{i+1}^{\prime}=0}^{p_{i+1}-1}M_{i}^{(b_{i},b_{i+1}^{\prime})}x_{i}^{p_{i}-1-b_{i}}x_{i+1}^{b_{i+1}^{\prime}}.
\]
Here, adjacent matrices share one common variable. The degree assignments
ensure proper alignment in the product polynomial
\begin{eqnarray*}
\tilde{M}^{(\text{MV}_{2})}(\mathbf{x}) & = & \prod_{i=0}^{m-1}\tilde{M_{i}}^{(\text{MV}_{2})}(x_{i},x_{i+1})
\end{eqnarray*}
Expanding the product yields
\begin{align}
\tilde{M}^{(\text{MV}_{2})}(\mathbf{x}) & =\sum_{\substack{b_{0}\in[p_{0}]\\
b_{m}^{\prime}\in[p_{m}]
}
}\sum_{\substack{\mathbf{b}\in\mathcal{N}_{1}^{m-1}\\
\mathbf{b}^{\prime}\in\mathcal{N}_{1}^{m-1}
}
}M^{(b_{0},\mathbf{b},\mathbf{b}^{\prime},b_{m}^{\prime})}\nonumber \\
 & \quad\times x_{0}^{p_{0}-1-b_{0}}x_{m}^{b_{m}^{\prime}}\prod_{i=1}^{m-1}x_{i}^{b_{i}^{\prime}+p_{i}-1-b_{i}}.\label{eq:product polynomial MV2}
\end{align}
In this case, the matrix blocks of the resultant matrix product $M^{(n_{0},n_{m})}$
for $\left(n_{0},n_{m}\right)\in[p_{0}]\times[p_{m}]$ can then be
obtain as the coefficient of the monomial
\[
x_{0}^{p_{0}-1-n_{0}}x_{m}^{n_{m}}\prod_{i=1}^{m-1}x_{i}^{p_{i}-1}
\]
which corresponds to $M^{(n_{0},n_{m})}=\sum_{\mathbf{n}\in\mathcal{N}_{1}^{m-1}}M^{(n_{0},\mathbf{n},\mathbf{n},n_{m})}.$

\textbf{Recovery threshold:} The product polynomial has degree $p_{0}-1$
in $x_{0}$, $p_{m}-1$ in $x_{m}$ and $2p_{i}-2$ in $x_{i}$ for
$i=1,\dots,m-1$. Thus,
\begin{equation}
R_{th}^{(\text{MV}_{2})}=p_{0}p_{m}\prod_{i=1}^{m-1}\left(2p_{i}-1\right).\label{eq:Rth MV2}
\end{equation}
This threshold is lower than multivariate Scheme 1, but still higher
than the univariate case.

\textbf{Storage capacity:} As in the previous case, let us first consider
the shared-memory setting. Here, the recovery threshold $R_{th}^{\text{(MV}_{2})}$
evaluations can be obtained by evaluating the polynomial over a Cartesian
product grid
\[
\mathbf{x}\in\mathcal{X}_{0}\times\mathcal{X}_{1}\times\dots\times\mathcal{X}_{m}
\]
with $\left|\mathcal{X}_{0}\right|=p_{0},$$\left|\mathcal{X}_{m}\right|=p_{m},$
and $\left|\mathcal{X}_{i}\right|=(2p_{i}+1)$ for $i=1,\dots,m-1$.
To enable these evaluations, workers store $\left|\mathcal{X}_{i}\right|\cdot\left|\mathcal{X}_{i+1}\right|$
coded block matrices of $\tilde{M_{i}}^{(\text{MV}_{2})}(x_{i},x_{i+1})$
corresponding to $\left(x_{i},x_{i+1}\right)\in\mathcal{X}_{i}\times\mathcal{X}_{i+1}$.
The number of coded block that need to be stored is 
\begin{eqnarray*}
S_{th,i}^{(\text{MV}_{2},\text{S})} & = & \begin{cases}
p_{0}(2p_{1}-1), & i=0,\\
(2p_{i}-1)(2p_{i+1}-1), & 1\leq i\leq m-2,\\
(2p_{m-1}-1)p_{m}, & i=m-1.
\end{cases}
\end{eqnarray*}
This requirement is higher than MV$_{1}$, but still lower than the
univariate scheme. Workers then compute the chain products of the
corresponding coded blocks. 

For the case, when workers have a dedicated memory. Again, consider
$N$ workers with homogeneous storage capacity for simplicity. In
this case, to every worker we assign $s_{0}s_{1}p_{0}(2p_{1}-1)$
and $s_{m}s_{m-1}p_{m}(2p_{m-1}-1)$ coded partitions of matrix $M_{0}$
and $M_{m-1}$, respectively, and $s_{i}s_{i+1}(2p_{i}-1)(2p_{i+1}-1)$
coded partitions of matrices $M_{i}$ for $i\in\left\{ 1,\dots m-2\right\} $.
The evaluation points for each worker form a disjoint Cartesian product
grid
\[
\mathbf{x}^{(w)}\in\mathcal{X}_{0}^{(w)}\times\mathcal{X}_{1}^{(w)}\times\dots\times\mathcal{X}_{m}^{(w)}
\]
with $\left|\mathcal{X}_{0}^{(w)}\right|=s_{0}p_{0},$$\left|\mathcal{X}_{m}^{(w)}\right|=s_{m}p_{m},$
and $\left|\mathcal{X}_{i}^{(w)}\right|=s_{i}(2p_{i}+1)$. The fractions
$1<s_{i}\leq1$ must be chosen so that the set size $\left|\mathcal{X}_{i}^{(w)}\right|$
is an integer. The $s_{i}(2p_{i}-1)s_{i+1}(2p_{i+1}-1)$ evaluation
points associated to $M_{i}$ are choose from $\mathcal{X}_{i}^{(w)}\times\mathcal{X}_{i+1}^{(w)}$.
Thus, for $i=1,\dots m-2$
\[
S_{th,i}^{(\text{MV}_{2},\text{D})}=Ns_{i}(2p_{i}-1)s_{i+1}(2p_{i+1}-1),
\]
and similarly $S_{th,0}^{(\text{MV}_{2},\text{D})}=Ns_{0}p_{0}s_{1}(2p_{1}-1)$
and $S_{th,M-1}^{(\text{MV}_{2},\text{D})}=Ns_{m}s_{m-1}p_{m}(2p_{m-1}-1)$
.

By computing all the chain products from the possible combinations
of coded block matrices locally available, that result on a evaluation
point in the Cartesian grid, the number of matrix chain products/evaluations
each worker can obtain is 
\begin{equation}
\prod_{i=0}^{m}\left|\mathcal{X}_{i}^{(w)}\right|=s_{0}p_{0}\left(\prod_{i=1}^{m-1}s_{i}(2p_{i}-1)\right)s_{m}p_{m}\label{eq:Evaluations per worker}
\end{equation}
Thus, to guarantee at least $R_{th}^{(\text{MV}_{2})}$ computations,
we require 

\begin{equation}
Ns_{0}p_{0}\left(\prod_{i=1}^{m-1}s_{i}(2p_{i}-1)\right)s_{m}p_{m}\geq R_{th}^{(\text{MV}_{2})}
\end{equation}
or, equivalently, substituting (\ref{eq:Rth MV2}) into (\ref{eq:Evaluations per worker})
$N\prod_{i=0}^{m}s_{i}\geq1$. 

Again to build intuition, choose $s_{i}=N^{-\frac{1}{m+1}}$ for all
$i\in[m]$, then 
\begin{eqnarray*}
s_{th,i}^{(\text{MV}_{2},\text{D})} & = & N^{\frac{m-1}{m+1}}\left(2p_{i}-1\right)\left(2p_{i+1}-1\right)
\end{eqnarray*}
and consider the representative case where all the matrices are partitioned
equally, i.e. $p_{i}=p$, for all $i$. In this setting, we have
\[
R_{th}^{(\text{MV}_{2})}=p^{2}\left(2p-1\right)^{m-1},
\]
\[
S_{u,i}^{(\text{MV}_{2},\text{S)}}=\left(2p-1\right)^{2},S_{u,i}^{(\text{MV}_{2},\text{D})}=N^{\frac{m-1}{m+1}}(2p-1)^{2}.
\]
Then the computation and communication overheads relative to the single
server setting are given by
\begin{eqnarray*}
\delta^{(\text{MV}_{2})} & = & \left(2-\frac{1}{p}\right)^{m-1}-1\sim2^{m-1}-1+\mathcal{O}\left(p^{-1}\right)\\
\delta_{u}^{(\text{MV}_{2},\text{S})} & = & \left(2-\frac{1}{p}\right)^{2}-1\sim2^{2}-1+\mathcal{O}\left(p^{-1}\right)\\
\delta_{u}^{(\text{MV}_{2},\text{D})} & = & N^{\frac{m-1}{m+1}}\left(2-\frac{1}{p}\right){}^{2}-1\sim2^{2}N^{\frac{m-1}{m+1}}-1+\mathcal{O}\left(p^{-1}\right)
\end{eqnarray*}
Unlike the univariate scheme, the storage requirements here does not
scale as $\mathcal{O}\left(p^{m-1}\right)$. Instead, it remains asymptotically
constant in the shared-memory case and grows only linearly with $N$
in the dedicated-memory case. The computation cost, similar to univariate
polynomial coding, is independent of $p$ but grows with the chain
length as $2^{m-1}.$

Overall, compared to the univariate scheme, multivariate (MV) strategies
significantly reduce the upload communication cost, at the expense
of a higher recovery threshold and, consequently, an increased number
of computations. Table \ref{tab:Computation-and-communication} summarizes
the asymptotic results obtained.%

\begin{table*}
\centering \caption{Computation overheads ($\delta$) and storage overheads for workers
with a common ($\delta_{s}^{\text{C}}$) and dedicated storage ($\delta_{s}^{\text{D}}$)
(asymptotic).\label{tab:Computation-and-communication}}
\begin{tabular}{lccc}
\hline 
\textbf{Scheme} & $\boldsymbol{\delta}$ & $\boldsymbol{\delta_{s}^{\text{C}}}$ & $\boldsymbol{\delta_{s}^{\text{D}}}$\tabularnewline
\hline 
UV & $\mathcal{O}(p^{-2})$ & $\mathcal{O}(p^{m-1})$ & $\mathcal{O}(p^{m-1})$\tabularnewline
MV$_{1}$ & $\mathcal{O}(p^{m-1})$ & $0$ & $N^{\tfrac{m-1}{m}}-1$\tabularnewline
MV$_{2}$ & $2^{m-1}-1+\mathcal{O}(p^{-1})$ & $2^{2}-1+\mathcal{O}(p^{-1})$ & $2^{2}N^{\tfrac{m-1}{m+1}}-1+\mathcal{O}(p^{-1})$\tabularnewline
\hline 
\end{tabular}
\end{table*}

\subsection{Note on the decodability of multivariate polynomial codes}

Unlike univariate polynomial interpolation, multivariate interpolation
cannot be guaranteed merely by selecting a number of distinct evaluation
points equal to the product of the largest monomial’s degrees plus
one. A rigorous analysis of multivariate decodability is omitted here;
see \cite{hasirciouglu2021JSAC,hasirciouglu2022JSAC} for details
on its application to coded computing. If evaluation points are chosen
independently at random, almost decodability is achievable, with success
probability increasing with the finite field size. In practice, however,
worker points are drawn from a Cartesian grid rather than independently,
so achieving decodability may require extra computations, which we
shall consider as additiona overhead.

\section{Numerical Evaluation\label{sec:Numerical-Evaluation}}

\begin{figure}
\centering \includegraphics[scale=0.5]{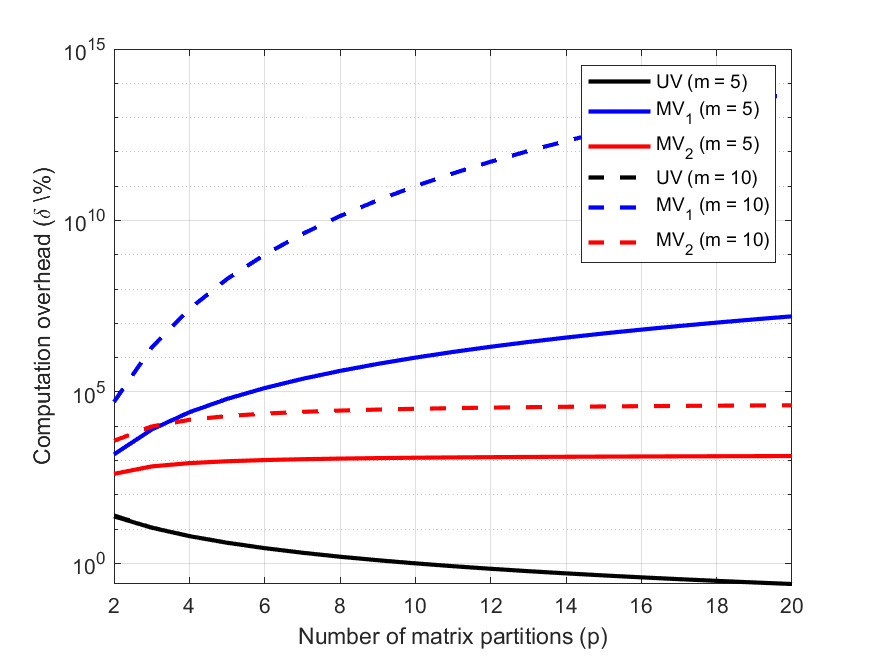}

\caption{Computation overhead, $\delta$\%, as a function of the number of
partitions.\label{fig:Computation-overhead}}
\end{figure}

\begin{figure}
\centering \includegraphics[scale=0.5]{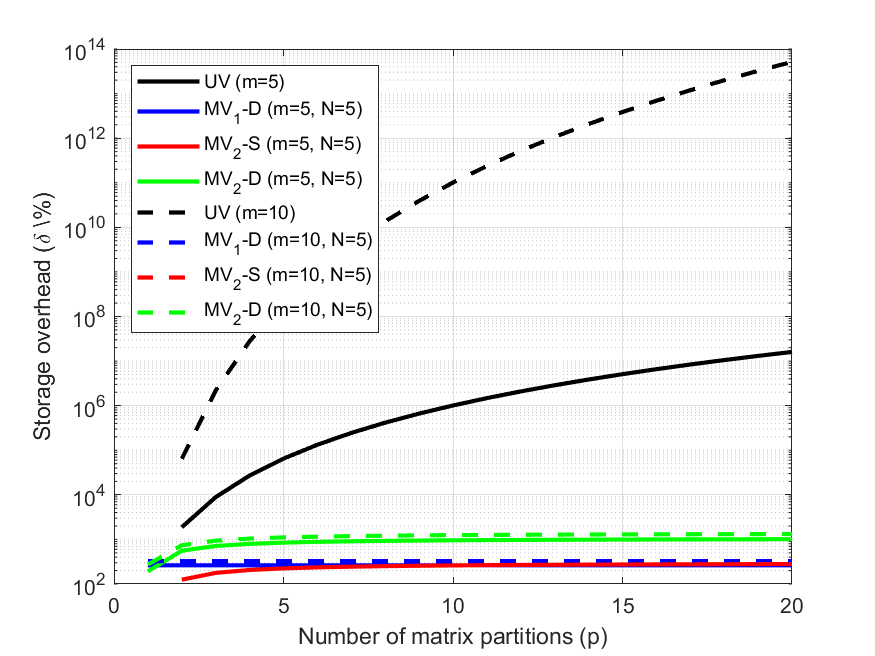}

\caption{Storage overhead, $\delta_{S}$\%, as a function of the number of
partitions.\label{fig:Storage-overhead}}
\end{figure}

\begin{figure}
\centering \includegraphics[scale=0.5]{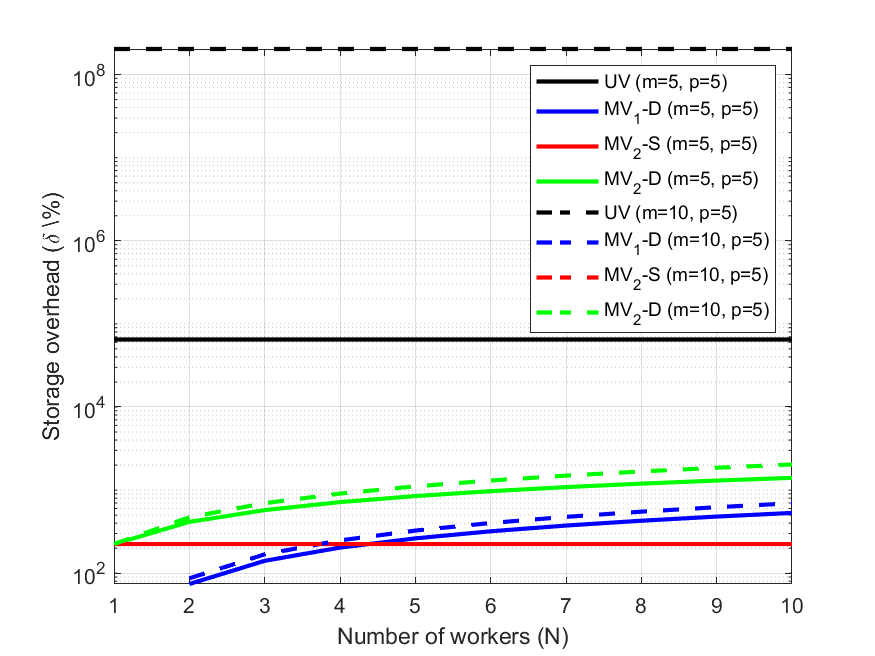}

\caption{Storage overhead, $\delta_{S}$\%, as a function of the number of
workers.\label{fig:Storage-overhead-1}}
\end{figure}

In Figures \ref{fig:Computation-overhead}, \ref{fig:Storage-overhead},
and \ref{fig:Storage-overhead-1} we evaluate the overhead expressions
obtained as a function the relevant parameters, $p$, $N$, and $m$.
Recall that we use logarithmic scale for the overheads which we represent
in percentage. For this evaluation we do not impose number of delivered
partitions/computations to be integers. Figure \ref{fig:Computation-overhead}
shows the computation overhead, defined as the percentage change in
element-wise computations required to recover the result relative
to a single-server baseline, as a function of the number of matrix
partitions (identical across all matrices in the chain). Results are
reported for matrix chains of ($m=5$) and ($m=10$). For the univariate
scheme (UV), the overhead decays rapidly with the number of partitions
as ($\mathcal{O}(p^{-2})$), independent of the chain length. In contrast,
for MV$_{1}$ the overhead grows quickly with both ($p$) and ($m$),
scaling as ($\mathcal{O}(p^{m-1})$), while for MV$_{2}$ it grows
only with ($m$), following ($2^{m-1}-1+\mathcal{O}(p^{-1})$). 

To illustrate the storage overhead as a function of the number of
partitions and workers, we consider two separate plots. Figure \ref{fig:Storage-overhead}
reports the storage overhead relative to a single-server baseline
as a function of the number of matrix partitions for a fixed number
of workers ($N=5$). Figure \ref{fig:Storage-overhead-1} shows the
overhead as a function of the number of workers for a fixed number
of partitions ($p=5$). Results are presented for ($m=5$) and ($m=10$).
For the univariate scheme (UV, black lines), the overhead is independent
of ($N$), but grows rapidly with both ($p$) and ($m$), scaling
as $(\mathcal{O}(p^{m-1}$)). For the multivariate (MV) schemes, we
distinguish between the dedicated and shared storage settings. In
all cases, the storage overhead converges quickly to a constant value,
which remains well below the UV curve: 0 for MV$_{1}$-S (omitted
since it cannot be displayed in logarithmic scale), ($2^{2}-1$) for
MV$_{1}$-D (blue), ($N-1$) for MV$_{2}$-S (red), and ($2^{2}N-1$)
for MV$_{2}$-D (green). 

\section{Conclusions\label{sec:Conclusions}}

We studied the problem of distributed matrix chain multiplication
under straggler-prone environments, extending coded computing techniques
beyond the conventional two-matrix setting. While univariate polynomial
codes effectively mitigate stragglers, their scalability is hindered
by rapidly increasing computational and storage costs as the chain
length grows.

To overcome these limitations, we introduced two multivariate polynomial
coding schemes that exploit higher-dimensional encodings to balance
computation and storage efficiency. Our analysis shows that, although
multivariate schemes introduce additional computation at the workers,
they significantly reduce storage and communication overheads. In
particular, the storage cost remains bounded with respect to the number
of matrices and partitions, in contrast to the exponential growth
observed in univariate approaches.

Overall, the results highlight a fundamental trade-off between computational
complexity and storage efficiency and demonstrate that the proposed
multivariate codes---especially MV$_{2}$---offer a practical and
scalable framework for large-scale distributed matrix chain multiplication.

\bibliographystyle{IEEEtran}
\bibliography{IEEEabrv,misRef}

\end{document}